\documentclass[10.5pt,compsoc]{tst}
\usepackage{graphicx}
\usepackage{footmisc}
\usepackage{subfigure}
\usepackage{url}
\usepackage{multirow}
\usepackage[noadjust]{cite}
\usepackage{amsmath,amsthm}
\usepackage{amssymb,amsfonts}
\usepackage{booktabs}
\usepackage{color}
\usepackage{ccaption}
\usepackage{booktabs}
\usepackage{float}
\usepackage{fancyhdr}
\usepackage{caption}
\usepackage{xcolor,stfloats}
\usepackage{comment}
\graphicspath{{figures/}}
\usepackage{cuted}  
\usepackage{captionhack}
\usepackage{epstopdf}
\usepackage{enumitem}
\usepackage{array}
\usepackage{stmaryrd}
\usepackage{algorithmic}
\usepackage[linesnumbered,vlined,ruled,nofillcomment]{algorithm2e}

\headevenname{\zihao{-5}{}}%
\headoddname{{\sf Shaowei Wang et al.:}\quad {\textbf{\emph{Segmented Private Data Aggregation
in the Multi-message Shuffle Model.}}}}%

\newtheoremstyle{mystyle}{0pt}{0pt}{\normalfont}{1em}{\bf}{}{1em}{}
\theoremstyle{mystyle}
\renewcommand\figurename{Fig.~}

\newtheorem{definition}{\textbf{Definition}}
\newtheorem{lemma}{\textbf{Lemma}}
\newtheorem{theorem}{\textbf{Theorem}}
\newtheorem{proposition}{\textbf{Proposition}}

\newcommand{\nop}[1]{}

\makeatletter
\renewcommand{\@biblabel}[1]{[#1]\hfill}
\makeatother

\begin{document}

\thispagestyle{plain}%
\thispagestyle{empty}%

\newcommand{\edit}[1]{{{#1}}}
\newcommand{\editg}[1]{{{#1}}}

\let\temp\footnote
\renewcommand \footnote[1]{\temp{\zihao{-5}#1}}
{}
\vspace*{-40pt}
\noindent{\zihao{5-}\textbf{\scalebox{0.885}[1.0]{\makebox[5.9cm][s]
{}}}}

\vskip .2mm
{\zihao{5-}
\textbf{
\hspace{-5mm}
\scalebox{1}[1.0]{\makebox[5.6cm][s]{}}}}

\vskip .2mm
{\zihao{5-}
\textbf{
\hspace{-5mm}
\scalebox{1}[1.0]{\makebox[5.6cm][s]{}}}}

\vskip .2mm\noindent
{\zihao{5-}\textbf{\scalebox{1}[1.0]{\makebox[5.6cm][s]{%
\color{white}{V\hfill o\hfill l\hfill u\hfill m\hfill%
e\hspace{0.356em}1,\hspace{0.356em}N\hfill u\hfill%
m\hfill b\hfill e\hfill r\hspace{0.356em}1,\hspace{0.356em}%
S\hfill e\hfill p\hfill t\hfill e\hfill%
m\hfill b\hfill e\hfil lr\hspace{0.356em}2\hfill0\hfill1\hfill8}}}}}\\

\begin{strip}
{\center
{\zihao{3}\textbf{
Segmented Private Data Aggregation\\ in the Multi-message Shuffle Model}}
\vskip 9mm}

{\center {\sf \zihao{5}
Shaowei Wang, Hongqiao Chen, Sufen Zeng, Ruilin Yang, Hui Jiang, Peigen Ye,  Kaiqi Yu, Rundong Mei, Shaozheng Huang, Wei Yang, and Bangzhou Xin
}
\vskip 5mm}
%

\centering{
\begin{tabular}{p{160mm}}

{\zihao{-5}
\linespread{1.6667} %
\noindent
\bf{Abstract:} {\sf
The shuffle model of differential privacy (DP) offers compelling privacy-utility trade-offs in decentralized settings (e.g., internet of things, mobile edge networks). Particularly, the multi-message shuffle model, where each user may contribute multiple messages, has shown that accuracy can approach that of the central model of DP. However, existing studies typically assume a uniform privacy protection level for all users, which may deter conservative users from participating and prevent liberal users from contributing more information, thereby reducing the overall data utility, such as the accuracy of aggregated statistics. In this work, we pioneer the study of segmented private data aggregation within the multi-message shuffle model of DP, introducing flexible privacy protection for users and enhanced utility for the aggregation server. Our framework not only protects users' data but also anonymizes their privacy level choices to prevent potential data leakage from these choices. To optimize the privacy-utility-communication trade-offs, we explore approximately optimal configurations for the number of blanket messages and conduct almost tight privacy amplification analyses within the shuffle model. Through extensive experiments, we demonstrate that our segmented multi-message shuffle framework achieves a reduction of about 50\% in estimation error compared to existing approaches, significantly enhancing both privacy and utility.}
}

\end{tabular}
}
\vskip 6mm

\vskip -3mm
\zihao{6}\end{strip}

\thispagestyle{plain}%
\thispagestyle{empty}%
\makeatother

\begin{figure}[b]
\vskip -6mm
\begin{tabular}{p{44mm}}
\toprule\\
\end{tabular}
\vskip -4.5mm
\noindent
\setlength{\tabcolsep}{1pt}
\begin{tabular}{p{1.5mm}p{79.5mm}}

$\bullet$& Shaowei Wang, Hongqiao Chen, Sufen Zeng, , Ruilin Yang, Peigen Ye, Kaiqi Yu, Rundong Mei, and Shaozheng Huang are with Institute of Artificial Intelligence, Guangzhou University,Guangzhou 510006, China. E-mail: wangsw@gzhu.edu.cn\\
$\bullet$& Hui Jiang is with China United Network Communications Group Corporation Limited, Beijing, China. E-mail: jiangh272@chinaunicom.cn\\
$\bullet$& Wei Yang is with Hefei National Laboratory, University of Science and Technology of China, Hefei 230026, China. E-mail: qubit@ustc.edu.cn\\
$\bullet$& Bangzhou Xin is with Institute of computer application, Chinese Academy of Engineering Physics, Mianyang, 621900, China. E-mail: xbw401@gmail.com
\end{tabular}
\end{figure}\zihao{5}

\zihao{5}
\noindent
\vspace{3.5mm}
\section{Introduction}
\label{s:introduction}
\noindent
User data is essential for numerous data-driven applications, including artificial intelligence of things, business intelligence, web usage mining, and medical data analysis. Privacy regulations, such as the GDPR in the EU, the CCPA in California, and the Personal Information Protection Law in China, now mandate rigid data privacy protections and give users the explicit right to opt into data collection processes. These regulations aim to safeguard user privacy and ensure that personal data is managed responsibly and transparently. Differential privacy (DP \cite{dwork2006differential}) has emerged as a standard for privacy protection, widely used in sectors like internet services. Given the potential untrustworthiness of servers, the local model of DP is commonly employed \cite{erlingsson2014rappor,ding2017collecting}, where data is anonymized on the user's device before being transmitted to the server. However, local DP requires injecting substantial noise into each user’s data to meet DP constraints, which significantly degrades the quality of data-driven applications. \\

\noindent
Recently, the shuffle model of differential privacy \cite{bittau2017prochlo,erlingsson2019amplification} has emerged as a promising paradigm, offering excellent privacy-utility trade-offs in decentralized settings. \edit{It has been deployed in multiple large-scale industry systems, such as Apple and Google's Exposure Notification
Privacy-preserving Analytics \cite{applegoogle} and Apple’s Samplable Anonymous Aggregation system \cite{talwar2024samplable}.}  This model introduces an intermediary, semi-trustable shuffler who mixes data from multiple users before sending it to the server, thereby amplifying privacy and improving data utility. The role of the shuffler can be played by anonymous channels, trusted hardware, or edge servers in mobile edge networks. Current research explores various data analysis tasks in both the single-message shuffle model (e.g., in \cite{cheu2019distributed,feldman2023stronger}), where each user contributes one message that satisfies local DP, and the multi-message shuffle model (e.g., in \cite{ghazi2021power,ghazi2021differentially,luo2022frequency,girgis2024multi}), where each user can contribute multiple messages. The single-message shuffle model exhibits significant utility gaps when compared to the multi-message model \cite{ghazi2021power}, which, in contrast, has the potential to achieve accuracy comparable to that of the central DP model. \\

\noindent
From the users' perspective, many might hesitate to contribute data under a fixed policy set by the server, especially if the server adopts a high privacy budget to enhance utility, as observed in Apple's differential privacy implementations \cite{tang2017apple}. Conversely, from the server's perspective, a uniform privacy policy can deter conservative users from participating and prevent liberal users from contributing more precise data. This discrepancy highlights the need for adaptable privacy policies that can cater to the varying comfort levels of different user groups. \\

\noindent
To address this issue, we aim to implement agnostic segmented privacy protection within the multi-message shuffle model of differential privacy (DP), which represents the most promising paradigm in decentralized settings lacking fully trustable parties. This approach allows each user to freely select one of many privacy levels without revealing their choice to the server (or other potential privacy adversaries). Compared to segmented or personalized privacy protection approaches found in the DP literature (e.g., in \cite{wang2015personalized,nie2018utility,liu2023echo}), our strategy encounters several unique challenges:
\begin{itemize}
    \item \emph{Agnostic of each user's privacy level.} Our approach aims to protect each user's privacy level option from the server, acknowledging that this option might reveal sensitive information about the user's preferences or status (e.g., younger users often opt for lower levels of privacy protection). Furthermore, privacy level choices may reflect the user's true values, with individuals holding minority opinions possibly favoring higher levels of protection. Unlike existing works \cite{wang2015personalized,nie2018utility}, where users' privacy levels are visible to the server, our model conceals these levels. This poses a significant challenge, as the server typically requires knowledge of each user's privacy parameters to compute accurate and unbiased estimators.  
    \item \emph{Analyses of privacy amplification on versatilly-randomized messages.} In the traditional shuffle model, where a uniform privacy policy is applied, all users employ the same randomizer. However, supporting segmented or personalized privacy necessitates the use of diverse randomizers to accommodate individual privacy budgets, diverging from the uniform model. The primary utility of the shuffle model arises from privacy amplification through shuffling \cite{erlingsson2019amplification}, where other users' messages can obfuscate any individual's data. Achieving tight privacy amplification with versatile randomizers in the single-message shuffle model can be challenging, as highlighted in \cite{liu2023echo}. This complexity increases when users contribute multiple messages to enhance utility. Developing and analyzing strategies in the multi-message shuffle model that maintain tight privacy amplification while utilizing individualized privacy budgets is a significant challenge. 
\end{itemize} 

\noindent
In this work, we introduce a framework for agnostic segmented privacy protection in the multi-message shuffle model of DP, designed to safeguard both the privacy of user data and the users' choices of privacy levels. The core concept involves aggregating privacy level choices and user data separately using the shuffle model, allowing the server to utilize aggregated (and potentially noised) privacy levels to compute accurate estimators. We decouple input-dependent messages and blanket messages, focusing on leveraging segmented privacy solely through input-dependent messages. This approach simplifies the analyses of shuffle privacy amplification by applying the data processing inequality of DP and avoids the complexities associated with versatile randomizers. Moreover, our method ensures almost tight privacy amplification, as worst-case input-dependent messages from other users in our framework cannot mimic the victim's message. Additionally, within this framework, we have developed a concrete protocol for set-valued data aggregation, a critical data mining task with broad applications in various domains, such as the internet of things \cite{tsai2013data} and medical data analyses, and extensively researched within the DP community \cite{qin2016heavy, wang2018privset, wang2020set, chen2011publishing, wang2023locally}. \\

\noindent
The contributions of this work are as follows:
\begin{itemize}
\item We initiate the study of segmented privacy protection in the multi-message shuffle model. This model is advantageous due to its favorable privacy-utility trade-offs in decentralized settings, and our approach of segmented privacy protection can further benefit both users and the server.
\item We propose a general framework for agnostic segmented privacy protection within the multi-message shuffle model, designed to safeguard both user data and privacy levels. We simplify the privacy amplification analysis in this specific context by eliminating the need to analyze input-dependent messages from other users.
\item We develop concrete protocols for set-valued data analyses, providing almost tight privacy amplification guarantees. Additionally, we numerically derive optimized hyperparameters, such as the number of blanket messages and Poisson sampling rates, to approximately minimize the estimation error.
\item Through extensive experiments, we demonstrate the effectiveness and efficiency of our protocols, achieving over a $50\%$ reduction in error compared to uniform privacy cases and personalized privacy protection in decentralized local/single-message shuffle models.
\end{itemize}

\noindent
The remainder of this work is organized as follows. Section \ref{sec:related} reviews related works on personalized privacy protection and the shuffle model of DP. Section \ref{sec:pre} provides preliminary knowledge. Section \ref{sec:framework} presents the general framework of agnostic segmented privacy protection in the multi-message shuffle model, including a simplified approach to privacy amplification analysis. Section \ref{sec:protocol} gives the concrete protocols for set-valued data aggregation. Section \ref{sec:exp} shows experimental results. Finally, Section \ref{sec:conclusion} concludes the paper.

\section{Related Works}
\label{sec:related}

\subsection{Shuffle Model of Differential Privacy}
 
\noindent
The original model of differential privacy \cite{dwork2006differential} necessitates trusted data curators to collect raw data from users and subsequently release noisy answers to statisticians. The local model of DP \cite{kasiviswanathan2011can} alleviates this requirement by allowing each user to sanitize their own data before sharing it with the server or analysts. However, this approach often results in lower data utility in the local model. Recently, the shuffle model of differential privacy \cite{erlingsson2019amplification} has emerged as a compelling approach to data privacy protection, combining the benefits of the classical central model (i.e., relatively high data utility) and the local model (i.e., minimal trust in other parties). In the shuffle model, an intermediary shuffler anonymizes and randomly permutes messages from a user population before forwarding them to the server (e.g., data analysts). The shuffle model of DP anonymizes user messages through an intermediate shuffler before sending them to a server for analytics. The model's foundation lies in privacy amplification analysis \cite{balle2019privacy,feldman2022hiding,wang2023unified}, ensuring the global privacy level of shuffled messages. Depending on the number of messages a user can send, the shuffle model can be categorized as either  single-message \cite{erlingsson2019amplification,balle2019privacy,feldman2022hiding} multi-message \cite{ghazi2021power,cheu2022differentially}. Essentially, multi-message shuffle DP protocols has significant higher utility potentials than single-message ones \cite{ghazi2021power}, and could often reach the central accuracy (i.e. similar utility to Gaussian and Laplace mechanism in the central of DP).  

\subsection{Segmented Privacy Preservation}
\noindent
Early works on personalized/segmented  privacy preserving \cite{xiao2006personalized} utilizes $k$-anonymity notion \cite{sweeney2002k}, where sensitive attributions are generalized. Latterly, within the more theoretically-founded privacy notion of differential privacy, \cite{jorgensen2015conservative} studies personalized privacy with the help of a trusted data curator. Wang \textit{et al.} \cite{wang2015personalized} initializes the study of the personalized privacy preservation in the local model of DP, where each user sanitizes categorical data locally and independently with personalized privacy parameters. Subsequent works further improve estimation utility via weighted summation \cite{nie2018utility}, and extend to other tasks (e.g., mean estimation \cite{xue2022mean}, mobile crowdsourcing \cite{wang2018personalized}, and federated learning \cite{yang2021federated}). On the other hand, the local model of DP is long be criticized of low utility, as every user must inject sufficient noises to satisfy DP by herself/himself. Recently, the shuffle model of DP, has emerged as a promising paradigm to better privacy-utility trade-offs, in decentralized settings without fully trustable third parties. A very recent work \cite{liu2023echo} studies personalized privacy in the single-message shuffle model for numerical value aggregation. This work, for the first time, explores personalized/segmented privacy preservation in the \emph{multi-message shuffle model}, so as to provide more flexible and improved privacy-utility trade-offs.

\section{Preliminaries}\label{sec:pre}
\noindent
A list of notations used throughout the paper can be found in Table \ref{tab:notations}.

\subsection{Differential Privacy and Its Segmented Version}

\begin{definition}[Hockey-stick divergence]\label{def:hsd}
The Hockey-stick divergence between two random variables $P$ and $Q$ is:
\[D_{e^{\epsilon}}(P \| Q)=\int\max\{0,P(x)-e^\epsilon Q(x)\}\mathrm{d}x,\]
where $P$ and $Q$ denote both the random variables and their probability density functions.
\end{definition} 

\vspace{\baselineskip}
\noindent 
Two variables $P$ and $Q$ are $(\epsilon, \delta)$-indistinguishable if $\max\{D_{e^\epsilon}
(P \| Q),\ D_{e^\epsilon}(Q \| P)\}\leq \delta$. For two datasets of equal size that differ only by a single individual's data, they are referred to as \emph{neighboring datasets}. The classical differential privacy limits the divergence of query results on all possible neighboring datasets (see Definition \ref{def:dp}). Similarly, in the local setting that accepts a single individual's data as input, the local $(\epsilon,\delta)$-differential privacy is presented in Definition \ref{def:ldp}. When $\delta=0$, the concept is abbreviated as $\epsilon$-LDP.

\begin{definition}[Differential privacy \cite{dwork2006differential}]\label{def:dp}
A protocol $\mathcal{R}:\mathbb{X}^n\mapsto \mathbb{Z}$ satisfies $(\epsilon,\delta)$-differential privacy if, for all neighboring datasets $X, X'\in \mathbb{X}^n$, $\mathcal{R}(X)$ and $\mathcal{R}(X')$ are $(\epsilon,\delta)$-indistinguishable.
\end{definition}
\begin{definition}[Local differential privacy \cite{kasiviswanathan2011can}]\label{def:ldp}
A protocol $\mathcal{R}:\mathbb{X}\mapsto \mathbb{Y}$ satisfies local $(\epsilon,\delta)$-differential privacy if, for all $x,x'\in \mathbb{X}$, $\mathcal{R}(x)$ and $\mathcal{R}(x')$ are $(\epsilon,\delta)$-indistinguishable.
\end{definition}

\noindent
\textbf{Group composition. } When two datasets differ at $s$ entries ($s\geq 1$), the privacy guarantee of a DP mechanism degrades. The degradation is captured in Lemma \ref{lemma:group}.

\begin{lemma}[Group Composition of DP \cite{dwork2006differential}]\label{lemma:group}
If a protocol $\mathcal{R}:\mathbb{X}^n\mapsto \mathbb{Z}$ satisfies $(\epsilon,\delta)$-differential privacy for all neighboring datasets, then for datasets $X, X'\in \mathbb{X}^n$ that differ at most $s$ entries, the $\mathcal{R}(X)$ and $\mathcal{R}(X')$ are $(s\cdot\epsilon, s\cdot e^{s\cdot \epsilon}\cdot\delta)$-indistinguishable.
\end{lemma}

\noindent
\textbf{Data processing inequality. } The data processing inequality is a key feature of distance measures (e.g., Hockey-stick divergence) used for data privacy. It asserts that the privacy guarantee cannot be weakened by further analysis of a mechanism's output.

\begin{definition}[Data processing inequality]\label{def:postprocess}
A distance measure $D:\Delta(\mathcal{S})\times\Delta(\mathcal{S})\to[0,\infty]$ on the space of probability distributions satisfies the data processing inequality if, for all distributions $P$ and $Q$ in $\Delta(\mathcal{S})$ and for all (possibly randomized) functions $g:\mathcal{S}\to\mathcal{S'}$, 
\[D(g(P)\|g(Q))\le D(P\|Q).\]
\end{definition}

\begin{table}
\centering
\caption{List of notations.}
\label{tab:notations}
\vspace{2em}
\renewcommand{\arraystretch}{1.0}

\begin{tabular}{c|l}
\hline
\bfseries Notation & \bfseries Description\\
\hline
$[i]$ & $\{1,2,...,i\}$ \\
$[i:j]$ & $\{i,i+1,...,j\}$ \\

$\llbracket inequality \rrbracket$ & Iverson bracket, equals 
1 if true, otherwise 0\\

\hline

$\mathcal{S}$ & the shuffling procedure\\
$\mathcal{R}$ & the randomization algorithm\\
$K$ & the number of privacy levels\\
$E$ & the list of privacy levels\\
$L$ & users' privacy segmentation options\\
$n$ & the number of users\\
$\epsilon_i$ & the privacy budget of user $i\in [n]$ ($\epsilon_i\in E$)\\
$d$ & the dimension of user's data \\
$\mathbb{X}$ & the domain of user data\\
$\mathbb{Y}$ & the domain of a sanitized message\\

\hline

\end{tabular}
\end{table}

\noindent
\textbf{Segmented differential privacy.} In segmented differential privacy, we assume there is a list of privacy options $E=\{E_1,E_2,\ldots,E_K\}\in \mathbb{R}^K$ that is increasing (i.e., $E_k\leq E_{k+1}$ for $k\in [K-1]$). Each user $i\in [n]$ can freely chooses an level $k_i$ from $[K]$, with the aim to preserve $(E_{k_i},\delta)$-differential privacy. We denote the relation of two neighboring dataset $X,X'\in \mathbb{X}^n$ that differ (only) at the $i$-th entry as $X\sim_{i} X'$. The segmented differential privacy for users in $[n]$ is defined as follows:

\begin{definition}[Segmented Differential Privacy]\label{def:segdp}
Given segmented options $E=\{E_1,E_2,\ldots,E_K\}\in \mathbb{R}^K$, user levels $L=\{k_1,\ldots,k_n\}\in [K]^n$, and $\delta\in [0,1]$, a protocol $\mathcal{R}:\mathbb{X}^n\mapsto \mathbb{Z}$ satisfies $(E,L,\delta)$-segmented differential privacy if, for all $i\in [n]$ and any neighboring datasets $X, X'\in \mathbb{X}^n$ that $X\sim_i X'$, $\mathcal{R}(X)$ and $\mathcal{R}(X')$ are $(E_{k_i},\delta)$-indistinguishable. 
\end{definition}

\vspace{\baselineskip}
\noindent
Note that the assumption of each user adopting the same level of failure probability $\delta$ is not truely necessary, one can easily extend to the scenario where each user has various failure probability requirements.


\noindent
\textbf{Segmented DP vs. Personalized DP.} Rather than permitting each user to freely choose arbitrary privacy levels (i.e., fully personalized DP), this work advocates for the implementation of segmented DP in the multi-message shuffle model. The primary reason is that determining appropriate randomization hyperparameters with shuffle privacy amplification for each privacy level often requires $\Omega(n)$ computational complexity. Consequently, fully personalized DP might lead to intolerable $\Omega(n^2)$ computational demands. Essentially, when privacy options are made fine-grained, segmented DP can approximate the fully personalized case.

\subsection{The Shuffle Model of Differential Privacy}
\noindent
Follow conventions in the shuffle model based on randomize-then-shuffle \cite{cheu2019distributed,balle2019privacy}, we define a multi-message protocol $\mathcal{P}$ to be a list of algorithms $\mathcal{P} = (\{\mathcal{R}_i\}_{i\in [n]}, \mathcal{A})$, where $\mathcal{R}_i: \mathbb{X} \to \mathbb{Y}^*$ is the local randomizer of user $i$, and $\mathcal{A}: \mathbb{Y}^n \to \mathbb{Z}$ the analyzer in the data collector's side. 
The overall protocol implements a mechanism $\mathcal{P} : \mathbb{X}^n \to \mathbb{Z}$ as follows.
Each user $i$ holds a data record $x_i$ and a local randomizer $\mathcal{R}_{i}$, then computes message(s) $Y_i = \mathcal{R}_i(x_i)$.
The messages $Y_1,...,Y_n$ are then shuffled and submitted to the analyzer. We write $\mathcal{S}(Y_1\cup \ldots\cup Y_n)$ to denote the random shuffling step, where $\mathcal{S} : \mathbb{Y}^* \to \mathbb{Y}^*$ is a \emph{shuffler} that applies a uniform-random permutation to its inputs.
In summary, the output of $\mathcal{P}(x_1, \ldots, x_n)$ is denoted by $\mathcal{A} \circ \mathcal{S} \circ \mathcal{R}_{[n]}(X) = \mathcal{A}(\mathcal{S}(\mathcal{R}_1(x_1)\cup \ldots\cup \mathcal{R}_n(x_n)))$. \\

\noindent
The shuffle model assumed that all parties involved in the protocol follow it faithfully and there is no collusion between them. From a privacy perspective, the goal is to ensure the differential privacy of the output $\mathcal{P}(x_1, \ldots, x_n)$ for any analyzer $\mathcal{A}$. By leveraging the post-processing property of the Hockey-stick divergence, it suffices to ensure that the shuffled messages $\mathcal{S} \circ \mathcal{R}_{[n]}(X)=\mathcal{S}(\mathcal{R}_1(x_1)\cup \ldots\cup \mathcal{R}_n(x_n))$ are differentially private. We formally define differential privacy in the shuffle model in Definition \ref{def:sdp}.

\begin{definition}[Differential Privacy in the Shuffle Model]\label{def:sdp}
A protocol $\mathcal{P} = (\{\mathcal{R}_i\}_{i\in [n]}, \mathcal{A})$ satisfies $(\epsilon,\delta)$-differential privacy in the shuffle model iff for all neighboring datasets $X, X'\in \mathbb{X}^n$, the $\mathcal{S} \circ \mathcal{R}_{[n]}(X)$ and $\mathcal{S} \circ \mathcal{R}_{[n]}(X')$ are $(\epsilon,\delta)$-indistinguishable.
\end{definition}

Similarly, we have the segmented differential privacy in the shuffle model as follows:
\begin{definition}[Segmented Differential privacy in the Shuffle Model]\label{def:segsdp}
Given segmented options $E=\{E_1,E_2,\ldots,E_K\}\in \mathbb{R}^K$, user levels $L=k_1,\ldots,k_n\in [K]^n$, and $\delta\in [0,1]$, a protocol $\mathcal{P} = (\{\mathcal{R}_i\}_{i\in [n]}, \mathcal{A})$ satisfies $(E,L,\delta)$-segmented differential privacy in the shuffle model if, for all $i\in [n]$ and any neighboring datasets $X, X'\in \mathbb{X}^n$ that $X\sim_i X'$, $\mathcal{S} \circ \mathcal{R}_{[n]}(X)$ and $\mathcal{S} \circ \mathcal{R}_{[n]}(X')$ are $(E_{k_i},\delta)$-indistinguishable.
\end{definition}

\subsection{Privacy Amplification via Shuffling}
\noindent
A key ingredient of the shuffle model is the privacy amplification via shuffling \cite{erlingsson2019amplification,cheu2019distributed}, which shows the privacy guarantee of $\mathcal{R}_{i}(x_i)$ can be amplified after shuffling. In this work, we adopt the state-of-the-art privacy amplification framework of variation-ratio reduction \cite{wang2023unified}. Considering two neighboring datasets $X$ and $X'$ that differ only in the $v$-th user's data (i.e. $X\sim_v X'$, $v\in [n]$, the $v$-th user is considered as the victim in DP), i.e., $X=\{x_1,\ldots,x_v,\ldots,x_n\}$ and $X'=\{x_1,\ldots,x'_v,\ldots,x_n\}$, where $x_1,\ldots,x_v,x'_v,\ldots,x_n\in \mathbb{X}$. One can define the following properties on (independent) local randomizers $\{{R}_i\}_{i\in [n]}$ with some parameters $p>1$, $\beta\in [0, \frac{p-1}{p+1}]$, and $q\geq 1$:
\begin{itemize}
\item[I.] \textit{$(p, \beta)$-variation property}: we say that the $(p, \beta)$-variation property holds if $D_{p}({R}_1(x_v)\| 
{R}_1(x'_v))=0$ and $D_{e^0}({R}_1(x_v)\|
{R}_1(x'_v))\leq \beta$ for all possible $x_v,x'_v\in \mathbb{X}$.
\item[II.] \textit{$q$-ratio property}: we say that the $q$-ratio property holds if $D_{q}({R}_v(x_v)\| {R}_i(x_i))=0$ and $D_{q}({R}_v(x'_v)\| {R}_i(x_i))=0$ hold for all possible $\{x_i\}_{i\in[n]\backslash\{v\}}\in \mathbb{X}^{n-1}$ and all $\{{R}_i\}_{i\in [n]\backslash\{v\}}$.
\end{itemize}

\noindent
Let $\mathcal{R}_{[n]}(X)$ denote the sanitized messages of $X$ by $\{{R}_i\}_{i\in [n]}$, the variation-ratio analyses \cite{wang2023unified} reduce the privacy amplification problem into a divergence bound between two counts (see Theorem \ref{the:reduction}).

\begin{theorem}[Variation-ratio reduction \cite{wang2023unified}]\label{the:reduction}
For $p> 1, \beta\in [0, \frac{p-1}{p+1}],  q\geq 1$, let $C\sim Binom(n-1, \frac{2\beta p}{(p-1)q})$, $A\sim Binom(C, 1/2)$ and $\Delta_1\sim Bernoulli(\frac{\beta p}{p-1})$ and $\Delta_2\sim Bernoulli(1-\Delta_1, \frac{\beta}{p-1-\beta p})$; let $P^{q,n}_{p,\beta}$ denote $(A+\Delta_1, C-A+\Delta_2)$ and $Q^{q,n}_{p,\beta}$ denote $(A+\Delta_2, C-A+\Delta_1)$.
If randomizers $\{{R}_i\}_{i\in [n]}$ satisfy the $(p, \beta)$-variation property and the $q$-ratio property for all possible $X,X'\in \mathbb{X}^n$ that $X\sim_v X'$ ($v\in [n]$), then for any measurement $D$ satisfying the data-processing inequality:
\begin{alignat*}{2}
&D(\mathcal{S}\circ\mathcal{R}_{[n]}(X)\|\mathcal{S}\circ\mathcal{R}_{[n]}(X'))
\leq D(P^{q,n}_{p,\beta}\|Q^{q,n}_{p,\beta}).
\end{alignat*}
\normalsize
\end{theorem}

\noindent
We call messages from other users as blanket messages for user $v$. \edit{Asymptotically, when $n$ is sufficiently large, by analyzing the Hockey-stick divergence and applying Hoeffding's tail bounds to the variables in Theorem \ref{the:reduction}, the variation-ratio reduction implies that the differential privacy level can be amplified to}: 
\begin{alignat}{1}\label{eq:nprivacy}
O\Big(\sqrt{\beta(p-1)q\log(1/\delta)/(p\cdot n)}\Big).
\end{alignat}
\edit{It shows that the amplification effect gets stronger when $\beta,p,q$ are small and the number of blanket messages $n-1$ getting large (see \cite{wang2023unified}[Section 4.1] for detail).} Besides asymptotic bounds, the variation-ratio reduction also provides tight numerical methods for Hockey-stick divergence computation and DP amplification. Our experiments will use these numerical amplification bounds for better privacy-utility trade-offs.

\section{A Personalized Multi-message Shuffle DP Framework}\label{sec:framework}
\noindent
In this part, we introduce the general data aggregation framework in the multi-message shuffle model with agnostic segmented privacy preservation.

\subsection{Trust Model}
\noindent
\edit{Following conventions in the literature, we assume that the shuffler is trustworthy or semi-trustworthy (if users encrypt each message with the server's public key), and that the shuffler does not collude with the server or the users. Our objective is to protect both users' data and their privacy-level options from the server, a concept termed \emph{agnostic segmented privacy protection}. Specifically, in the shuffle model considered in this work, the server (i.e., a potential privacy adversary) observes only shuffled messages (regarding privacy levels or sanitized user data) from the shuffler. In our framework design, the server cannot observe the association between a user and the user's chosen privacy option through shuffle-based anonymization and possible differential privacy. For user data, we aim to preserve segmented differential privacy (see Definition \ref{def:segsdp}) against privacy adversaries, including the server.
}

\subsection{Design Goals}
\noindent
The design goals of our framework are as follows:
\begin{itemize}
    \item \textbf{Personalized protection of users' data. } The fundamental goal of our framework is preventing adversaries (e.g., the server) from inferring each user's data, at the user-specified level of differential privacy.
    \item \textbf{Protection of users' choices of privacy levels. } Privacy adversaries should not have access to each user's privacy level choice, as it relates to the preference/status of the user (e.g., younger users often adopt lower level of privacy protection). Besides, the privacy level choice may also relate to the true value the user holds, such as users with minority values might tend to use higher level of privacy protection.
    \item \textbf{High aggregation utility. } Our framework aims to not only provide flexible privacy protection for users, but also give improved utility for the server. Though the multi-message shuffle model inherently have more utility potential than the single-message shuffle model and the local model, imposing protection of user's privacy level can harm the accuracy as the server has no exact information about each user's randomization parameters, and coordinating users to provide personalized DP might also sacrifice utility.
\end{itemize}


\subsection{Overall Procedures}

\begin{figure}[tb]
\begin{center}
\centerline{\includegraphics[width=80mm]{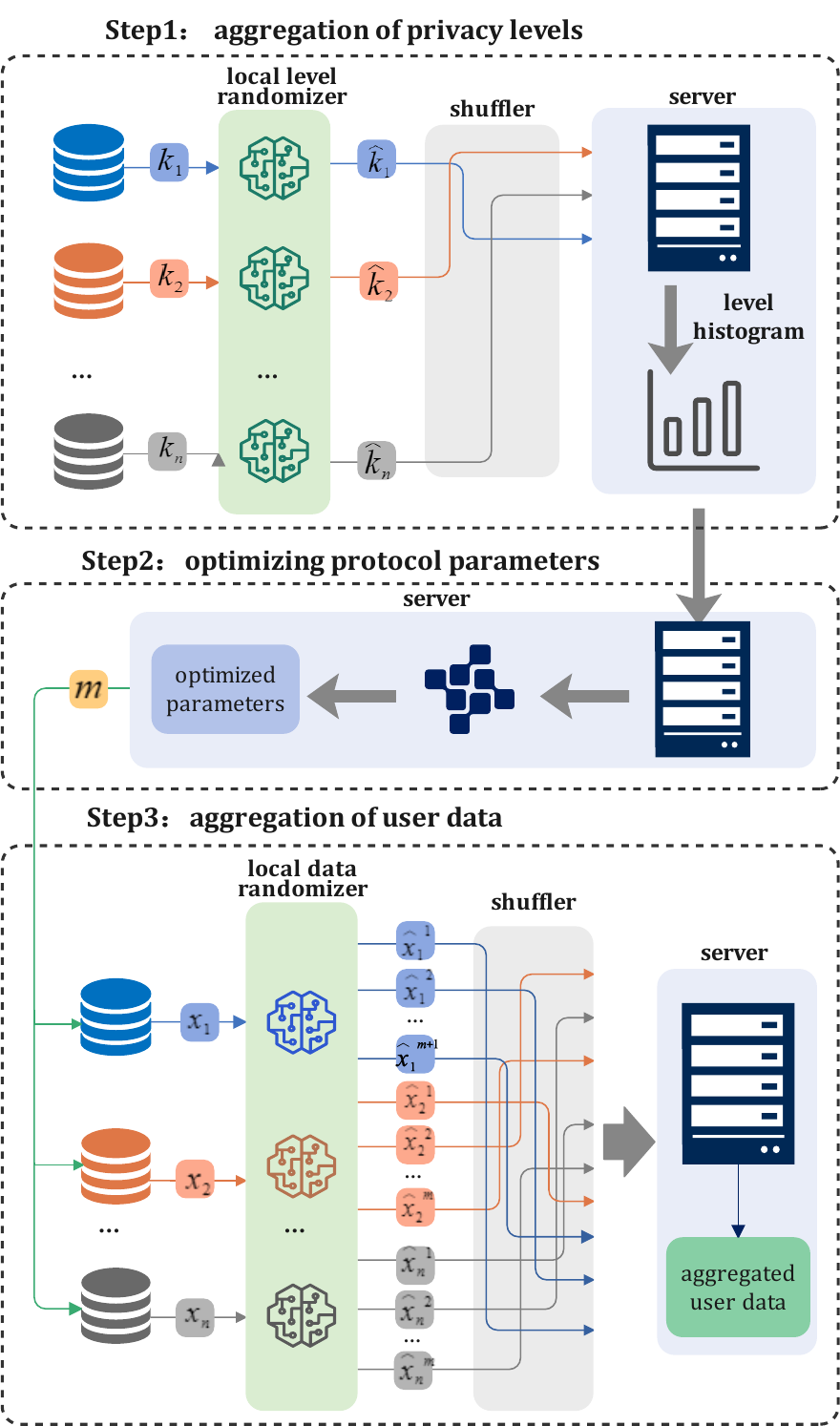}}
\vspace*{-0.5em}
\caption{Illustration of multi-message shuffle DP framework with agnostic segmented privacy preservation.}
\label{fig:framework}
\end{center}
\vspace*{-1.2em}
\end{figure}

\noindent
In this section, we describe the major steps in our personalized multi-message shuffle DP framework (see Figure \ref{fig:framework} for illustration). To protect the privacy level choices of users (i.e., the second goal), we separately collect privacy levels and data from users, allowing the shuffler to shuffle users' privacy levels. To achieve high aggregation utility (i.e., the third goal), the server calibrates optimized randomization/blanket strategies after receiving the shuffled privacy levels from users, with the objective of minimizing expected aggregation error while protecting personalized DP. Users then follow the optimized randomization/blanket configuration and send the generated messages to the shuffler. Finally, the server receives shuffled messages from the shuffler and derives the (unbiased) aggregation result.

\begin{algorithm}[h]
\caption{A Multi-message Shuffle DP Framework with Segmented Privacy Preservation} \label{alg:framework}
\SetKwInOut{Parameter}{Params}
\Parameter{Users' data $\{x_1,\ldots,x_n\}\in \mathbb{X}$, privacy levels $E=\{E_1,\ldots,E_n\}$, users' privacy options $L=\{k_1,\ldots,k_n\}\in E^n$, DP failure parameter $\delta\in [0,1]$, algorithms $\mathcal{R}_p$, $\mathcal{S}_p$ and $\mathcal{A}_p$ for privacy level aggregation, algorithms $optimizeParamter$, $\mathcal{S}$ and $\mathcal{A}$ for data aggregation.}

\setlength{\algomargin}{4em} 

\tcc{step 1: aggregation of privacy levels}

\For{user $i\in [n]$}{


    {$P_{i}\leftarrow \mathcal{R}_p(k_i)$}

    {send messages in $P_{i}$ to the shuffler}
    
}

\tcc{on the shuffler side, shuffles messages about privacy levels}

{$M_p\leftarrow \mathcal{S}_p(P_{1}\cup P_{2}\ldots \cup P_{n})$}

\tcc{on the server side, obtain (estimated) histogram of privacy levels}

{$\{\hat{n}_1,\ldots,\hat{n}_K\}\leftarrow \mathcal{A}_p(M_p)$}

\tcc{step 2: optimizing protocol parameters, on the server side}

{$m,\mathcal{R}_{blanket},\mathcal{R}_{(1)},\ldots,\mathcal{R}_{(K)} \leftarrow optimizeParameter(E,\delta,\{\hat{n}_1,\ldots,\hat{n}_K\})$}

\tcc{step 3: aggregation of user data}

\For{user $i\in [n]$}{

    {$M_i\leftarrow \mathcal{R}_{k_i}(x_i)\cup \mathcal{R}_{blanket}(m)$}

    {send messages in $M_{i}$ to the shuffler}
    
}

\tcc{on the shuffler side, shuffles messages about user data}

{$M\leftarrow \mathcal{S}(M_{1}\cup M_{2}\ldots \cup M_{n})$}

\tcc{on the server side, obtain aggregation results about user data}

{$A\leftarrow\mathcal{A}(M,M_p)$}

\end{algorithm}

\noindent
\textbf{Step 1: Aggregation of privacy levels. } Without loss of generality, we assume there are $K$ options of DP level $E=\{E_1,E_2,\ldots,E_K\}\in \mathbb{R}^K$, and each user's choice $k_i$ belongs to $[K]$. Let $n_k$ denote the number of users that choose $E_k$: $n_k=\#\{i\ |\  i\in [n]\ \text{and}\ k_i=k\}$. The first step of our framework will obtain a histogram $\{n_1,\ldots, n_k\}$ or its noisy estimator, through the shuffle model without (or with) local randomization. In case we only need to anonymize privacy levels from users, each user $i$ can simply send $k_i$ to the shuffler; in case more rigid protection on privacy levels is needed, the users can invoke a shuffle protocol $\mathcal{P}=\mathcal{A}_{p}\circ\mathcal{S}_p\circ\mathcal{R}_p$ to obtain a $\epsilon_p$-differentially private histogram, which is well studied in the literature. We denote the observed count of level $n_k$ as $\hat{n}_k$, either the exact values or estimators. 

\noindent
\textbf{Step 2: Optimizing protocol parameters. } Given the observed level counts $\hat{n}_1,\ldots, \hat{n}_K$, the server derives an optimized randomization parameters $s$, which often include the number of blanket messages $m$ per user and the local randomizer $\mathcal{R}_{k}$ for each privacy level $k$. The parameters optimization objective is pertained to the aggregation task, such as being minimizing the (upper bound) of mean squared error of aggregation results about users' data $\{x_1,\ldots,x_n\}$. In the the shuffle model, the number of blanket messages have critical effects on the aggregation error and the privacy guarantee. Besides the number of blanket message, the local randomizer $\mathcal{R}_{k}$ also affect the privacy guarantee of the local data $x_i$ (where $\epsilon_i=E_k$), it has to be carefully designed to ensure the amplified privacy confrom $(E_k, \delta)$-DP.

\noindent
\textbf{Step 3: Aggregation of user data. } After receiving the number of blanket messages $m$ per user and the local randomizer $\mathcal{R}_{k}$ for each privacy level $k$, each user $i$ with $\epsilon_i=E_k$ uses the local randomizer $\mathcal{R}_{k}$ on $x_i$ to generate zero or one input-dependent message. Each user also generates multiple blanket messages with $\mathcal{R}_{blanket}(m)$, which will general $m$ input-independent messages in expectation, where the $m\in \mathbb{R}^{\geq 0}$ can be a real-valued number. These messages are then sent to the shuffler. The shuffler runs the uniform permutation procedure $\mathcal{S}$ on all received messages $\mathcal{R}_{k_1}(x_1)\cup\mathcal{R}_{blanket}(m)\cup \mathcal{R}_{k_2}(x_2)\cup\mathcal{R}_{blanket}(m)\cup\ldots\cup\mathcal{R}_{k_n}(x_n)\cup \mathcal{R}_{blanket}(m)$, and send them to the server. Upon receiving shuffled messages $\mathcal{S}(\mathcal{R}_{k_1}(x_1)\cup \mathcal{R}_{blanket}(m)\cup\ldots\cup\mathcal{R}_{k_n}(x_n)\cup \mathcal{R}_{blanket}(m))$, the server runs the aggregation algorithm $\mathcal{A}$ on them to obtain the aggregation results.

\subsection{Simplification of Privacy Guarantee}
\noindent
In the user data aggregation step, each user generates input-dependent message(s) via $\mathcal{R}_{k_i}(x_i)$ and $m$ input-independent message(s) via $\mathcal{R}_{blanket}(m)$. In this section, we demonstrate that the privacy guarantee for each user can be simplified by ignoring $\mathcal{R}_{k{i'}}(x_{i'})$ for all $i' \in [n] \backslash {i}$. This simplification facilitates the subsequent privacy amplification procedures.

\begin{lemma}[Simplification of Privacy Guarantee]\label{lemma:simplification}
With the same privacy options $L=\{k_1,\ldots,k_n\}$, for neighboring dataset $X=\{x_1,\ldots,x_n\}\in \mathbb{X}^n$ and $X'=\{x'_1,\ldots,x'_n\}\in\mathbb{X}^n$ that differ at the $i$-th user's data, let $M_i$ denote the messages of $\mathcal{R}_{k_i}(x_i)\cup\mathcal{R}_{blanket}(m)$ and $M'_i$ denote the messages of $\mathcal{R}_{k_i}(x'_i)\cup\mathcal{R}_{blanket}(m)$. Then, for any distance measure $D$ that satisfy the data processing inequality, we have:
\begin{alignat*}{2}
&D\big(\mathcal{S}(M_{1}\cup M_{2}\ldots \cup M_{n})\|\mathcal{S}(M'_{1}\cup M'_{2}\ldots \cup M'_{n})\big)\\
\leq&D\big(\mathcal{S}(\mathcal{R}_{k_i}(x_i)\cup \mathcal{R}_{blanket}(m)\ldots \cup \mathcal{R}_{blanket}(m))\\
&\ \ \ \ \|\mathcal{S}(\mathcal{R}_{k_i}(x'_i)\cup \mathcal{R}_{blanket}(m)\ldots \cup \mathcal{R}_{blanket}(m))\big).
\end{alignat*}
\end{lemma}

\begin{IEEEproof}
To prove this result, we use the data processing inequality of the distance measure $D$. Specifically, Given $\mathcal{S}(\mathcal{R}_{k_i}(x_i)\cup \mathcal{R}_{blanket}(m)\ldots \cup \mathcal{R}_{blanket}(m))$ or $\mathcal{S}(\mathcal{R}_{k_i}(x'_i)\cup \mathcal{R}_{blanket}(m)\ldots \cup \mathcal{R}_{blanket}(m))$ as the input $I$, we design the following two-step post-processing function:
\begin{itemize}
    \item[(1)] For each $i'\in [n]\backslash\{i\}$, generate input-dependent message $\mathcal{R}_{k_{i'}}(x_{i'})$. Note that here $x_{i'}=x'_{i'}$.
    \item[(2)] Uniform-randomly permute $I\cup\{\mathcal{R}_{k_{i'}}(x_{i'})\}_{i'\in [n]\backslash\{i\}}$ to obtain output $O$.
\end{itemize}
When the input $I$ is $\mathcal{S}(\mathcal{R}_{k_i}(x_i)\cup \mathcal{R}_{blanket}(m)\ldots \cup \mathcal{R}_{blanket}(m))$, we have the output $O$ distributionally equals to $\mathcal{S}(M_{1}\cup M_{2}\ldots \cup M_{n})$; when the input $I$ is $\mathcal{S}(\mathcal{R}_{k_i}(x'_i)\cup \mathcal{R}_{blanket}(m)\ldots \cup \mathcal{R}_{blanket}(m))$, we have the output $O$ distributionally equals to $\mathcal{S}(M'_{1}\cup M'_{2}\ldots \cup M'_{n})$. Therefore, according to the data processing inequality, we have the conclusion.
\end{IEEEproof}

\vspace{\baselineskip}
\noindent
The Lemma \ref{lemma:simplification} demonstrate that we can focus on $\mathcal{R}_{k_i}$ and the input-independent blanket messages provided by other users to obtain privacy guarantee of each user $i$. This is necessary for the personalized DP in the multi-message shuffle model, as the input-dependent messages from other users can be quite heterogeneous rendering the privacy analyses extremely complicated. In contrast, the simplified situation is more homogeneous.

\vspace{\baselineskip}
\noindent
In our subsequent concrete protocol for set-valued data analyses, the input-dependent messages from other users can be noiseless, thus providing no blanket or amplification effects. In this case, the simplification here is actually tight—i.e., there are instances where users $[n] \backslash \{i\}$ hold neither elements in $x_i^0$ nor those in $x_i^0$ (see Section \ref{subsec:privacy}).

\section{A Protocol for Set-valued Data Analyses}\label{sec:protocol}
\noindent
In this section, within the proposed framework, we present a concrete segmented shuffle DP protocol for set-valued distribution analyses, a prevalent task in user data aggregation and mining. \\

\noindent
We assume each user holds a subset $x_i$ of the item domain $T$, where $T = \{T_1, \ldots, T_d\}$ and $|T| = d$. Without loss of generality, we assume $x_i$ contains exactly $s$ elements. That is, let $\mathbb{T}^s = \{x \mid x \subseteq T \text{ and } |x| = s\}$, then $x_i \in \mathbb{T}^s$. We assume each $x_i$ is independently and identically distributed (i.i.d.) and sampled from an unknown distribution $P_{\mathbb{T}^s}$. The aim is to estimate the distribution of each element:
$$
W = \{w_1, \ldots, w_d\} = \sum_{x \in \mathbb{T}^s} x \cdot P_{\mathbb{T}^s}(x),
$$
where $x$ is represented in binary vector form (i.e., $x \in \{0,1\}^d$, where the $j$-th element indicates the presence of the $j$-th element $T_j$ in $T$). \\

\noindent
To minimize the $q$ parameter in variation-ratio analyses and maximize privacy amplification, we use uniformly random elements over $T$ as blanket messages. Specifically, each blanket message is generated by uniformly and randomly selecting one element from the domain $T$. The input-dependent randomizer $\mathcal{R}_{k}$ is defined as probabilistically reporting each element $t$ with probability $\lambda_k$ (termed the \emph{Poisson randomizer}):
\begin{equation}\label{eq:r}
    y =
    \begin{cases}
    t, & \text{with probability } \lambda_{k}, \\
    \Phi, & \text{otherwise. }
    \end{cases}
\end{equation}

\begin{algorithm}[!htp]
    \caption{A Personalized Multi-message Shuffle DP Framework for Set-valued Data Analyses} \label{alg:set}
    
    \SetKwInOut{Parameter}{Params}
    \Parameter{Users' data $\{x_1,\ldots,x_n\}\in \mathbb{T}^s$, privacy levels $E=\{E_1,\ldots,E_n\}$, users' privacy options $L=\{k_1,\ldots,k_n\}\in E^n$, DP failure parameter $\delta\in [0,1]$, algorithms $\mathcal{R}_p$, $\mathcal{S}_p$ and $\mathcal{A}_p$ for privacy level aggregation, algorithms $optimizeParamter$, $\mathcal{S}$ and $\mathcal{A}$ for data aggregation.}
    
    \tcc{step 1: aggregation of privacy levels}
    
    \For{user $i\in [n]$}{
    
    
        {$P_{i}\leftarrow \mathcal{R}_p(k_i)$}
    
        {send messages in $P_{i}$ to the shuffler}
        
    }
    
    
    {$M_p\leftarrow \mathcal{S}_p(P_{1}\cup P_{2}\ldots \cup P_{n})$}
    
    
    {$\{\hat{n}_1,\ldots,\hat{n}_K\}\leftarrow \mathcal{A}_p(M_p)$}

    \tcc{step 2: optimizing protocol parameters, on the server side}

    {$m,\lambda_{1},\ldots,\lambda_{k} \leftarrow optimizeParameter(E,\delta,\{\hat{n}_1,\ldots,\hat{n}_K\})$}

    \tcc{step 3: aggregation of user data}
    
    \For{user $i\in [n]$}{

        {$M_i\leftarrow \Phi$}
    
        \For{element $t$ in $x_{i}$}{
    
            {sample $r\in uniform([0.0,1.0))$}
    
            \If{$r\leq \lambda_{(k_i)}$}{
                {$M_i\leftarrow M_i\cup t$}
            }
        }
    
        \For{$a$ in $[\lceil m\rceil]$}{
    
            {sample $r\in uniform([0.0,1.0))$}
    
            \If{$r\leq m/\lceil m\rceil$}{
                {$M_i\leftarrow M_i\cup uniform(T)$}
            }
        
        }
    
        {send messages in $M_{i}$ to the shuffler}
        
    }

    
    {$M\leftarrow \mathcal{S}(M_{1}\cup M_{2}\ldots \cup M_{n})$}
    
    
    \For{$j\in [d]$}{
        {$C_j\leftarrow \#\{y\ |\ y\in M\text{ and } y=T_j\}$}
    
        $\hat{W}_j\leftarrow \frac{C_j-{n\cdot m}/{d}}{\sum_{k\in [K]}\hat{n}_k\cdot \lambda_{k}}$
    
    }
    
    {return $\{\hat{W}_1,\ldots,\hat{W}_d\}$}
    
\end{algorithm}

\noindent
On the server side, the protocol parameters (e.g., $\lambda_{1}, \ldots, \lambda_{k}$ and the number of blanket messages $m$) are chosen with the objective of minimizing the error of the distribution estimator, the formula for which will be shown later in Section \ref{subsec:hyperparameter}. Regarding the distribution estimator, recall that the original unbiased estimator is given by:
$$
\widehat{\llbracket t=T_j\rrbracket} = \frac{\llbracket y=T_j\rrbracket}{\lambda_{k}}.
$$
In our framework, which protects privacy levels, a message $y$ and its corresponding privacy level $E_k$ (or its local parameter $\lambda_{k}$) are disassociated. Therefore, we estimate a weighted distribution (of item $j \in [d]$) that is still unbiased if $n_1,\ldots,n_K$ are exact:
$$
\mathbb{E}\left[\frac{\sum_{i \in [n]} \llbracket T_j \in x_i\rrbracket \cdot \lambda_{k}}{\sum_{i \in [n]} n_k \cdot \lambda_{k}}\right]=w_j.
$$
In case $n_1,\ldots,n_K$ are noised by DP (often being unbiased), the denominator can be unbiasedly estimated as $\sum_{k \in [K]} \hat{n}_k \cdot \lambda_{k}$, and the numerator can be unbiasedly estimated as $\sum_{y \in M} (\llbracket y = T_j\rrbracket) - \frac{n \cdot m}{d}$. Here, $\sum_{y \in M} (\llbracket y = T_j\rrbracket)$ is the number of $T_j$ in the shuffled messages $M$, and $\frac{n \cdot m}{d}$ is the expected count of $T_j$ contributed by blanket messages. The overall protocol is summarized in Algorithm \ref{alg:set}.

\subsection{Privacy Guarantees}
\noindent
In this section, we present the privacy guarantee of the proposed protocol for set-valued data. To achieve this, we first extend the variation-ratio privacy amplification to cases where the number of blanket messages is random (specifically, following a binomial distribution, as described in lines 13-16 of Algorithm \ref{alg:set}). Since each user might contribute multiple input-dependent messages, as described in lines 9-12 of Algorithm \ref{alg:set}, we then apply the group privacy composition of Differential Privacy (DP) to derive the final privacy guarantee for each user.

\subsubsection{Privacy Amplification with Random Number of Blanket Messages} 
\noindent
To balance privacy and utility in a fine-grained manner, every user might only contribute less than one blanket message in expectation (e.g., when $n$ is relatively large, as demonstrated in \cite{ghazi2021differentially, luo2022frequency} and in Section \ref{sec:exp}). Our protocol allows users to probabilistically contribute blanket messages, see lines 13-16 in Algorithm \ref{alg:set}. As a consequence, the original variation-ratio reduction that handles only a fixed number of blanket messages becomes inapplicable. In this part, we extend the privacy amplification analysis to a binomial random number of blanket messages.  \\

\noindent
We consider the case where each user holds one item (i.e., $s=1$), and define the following 4-step procedure $F_{rb}$ that takes as input $x$: (1) randomize $x$ with randomizer $\mathcal{R}_{k}$ to generate one message $\mathcal{R}_{k}(x)$, (2) sample $N \sim \text{Binomial}(n, \gamma)$ with some $p \in [0.0, 1.0]$, (3) generate blanket messages with $N$ independent calls of $\mathcal{R}_{blanket}(1)$, (4) uniformly randomly shuffle these messages to obtain the result $F_{rb}(x)$. We aim to analyze the differential privacy property of function $F$ with respect to input $x \in \mathbb{X}$, and give Proposition \ref{the:randomreduction}.

\begin{theorem}[Variation-ratio reduction with random-number blanket messages]\label{the:randomreduction}
If the randomizer $\mathcal{R}_{k}$ satisfy the $(p, \beta)$-variation property and each blanket message in the function $F_{rb}$ satisfy the $q$-ratio property w.r.t. to $\mathcal{R}_{k}$, then for any $x_1^0,x_1^1\in \mathbb{X}$ and any measurement $D$ satisfying the data-processing inequality:
\begin{alignat*}{2}
&D(F_{rb}(x_1^0)\|F_{rb}(x_1^1)) \leq D(P^{q/\gamma,n+1}_{p,\beta}\|Q^{q/\gamma,n+1}_{p,\beta}).
\end{alignat*}
\normalsize
\end{theorem}
\begin{IEEEproof}
For $n\in\mathbb{R}^{\geq 0}, \gamma\in [0.0,1.0], p> 1, \beta\in [0, \frac{p-1}{p+1}],  q\geq 1$, let $C\sim Binom(n, \gamma\cdot\frac{2\beta p}{(p-1)q})$, $A\sim Binom(C, 1/2)$ and $\Delta_1\sim Bernoulli(\frac{\beta p}{p-1})$ and $\Delta_2\sim Bernoulli(1-\Delta_1, \frac{\beta}{p-1-\beta p})$; let $U^{q,\gamma}_{p,\beta}$ denote $(A+\Delta_1, C-A+\Delta_2)$ and $V^{q,\gamma}_{p,\beta}$ denote $(A+\Delta_2, C-A+\Delta_1)$. Follow the same proof steps as the variation-ratio reduction's, and use the fact that: if $N\sim Binom(n, \gamma)$ and $C\sim Binom(N, \frac{2\beta p}{(p-1)q})$ then $C\sim Binom(n, \gamma\cdot\frac{2\beta p}{(p-1)q})$, we have $D(F(x_1^0)\|F(x_1^1)) \leq D(U^{q,\gamma}_{p,\beta}\|V^{q,\gamma}_{p,\beta}$. Now observe that  $U^{q,\gamma}_{p,\beta}=P^{q/\gamma,n+1}_{p,\beta}$ and $V^{q,\gamma}_{p,\beta}=Q^{q/\gamma,n+1}_{p,\beta}$, we have the conclusion.
\end{IEEEproof} 

\vspace{\baselineskip}
\noindent
The theorem indicates that the Bernoulli sampling at line 15 is equivalent to dividing the ratio parameter $q$ by a factor of the sampling rate $m / \lceil m \rceil$. Overall, there are $n$ calls of $\mathcal{R}_{blanket}(m)$ in Algorithm \ref{alg:set}, which is equivalent to sampling $N \sim \text{Binomial}(n \cdot \lceil m \rceil, m / \lceil m \rceil)$, and then generating $N$ blanket messages that are uniformly random over domain $T$. Though the analysis is conducted with $s=1$, the case $s>1$ can easily be derived using the group privacy property of DP.

\subsubsection{Privacy Amplification with Poisson Randomizer}
\noindent
To fully separate noise and information in the shuffle model, we use the Poisson randomizer in Equation \ref{eq:r} for each element $t$ in the user's input data $x_i$. In this part, we show that the Poisson randomizer reduces the total variation $\beta$ of the local randomizer to $\lambda_{k}$. \\

\noindent
Define the following 4-step procedure $H$ that takes as input $x$: (1) randomize $x$ with the Poisson randomizer of parameter $\lambda_{k}$, (2) sample $N \sim \text{Binomial}(n, \gamma)$ with some $\gamma \in [0.0, 1.0]$, (3) generate $N$ blanket messages that are uniformly random over domain $T$, (4) uniformly shuffle all these messages to obtain the result. We aim to analyze the differential privacy property of function $H$ with respect to input $x \in \mathbb{T}$, and provide Proposition \ref{the:randomreduction}.

\begin{lemma}[Variation-ratio reduction with Poisson randomizer]\label{lemma:poissonreduction}
\editg{For any $x_1^0,x_1^1\in \mathbb{T}$ and any measurement $D$ satisfying the data-processing inequality, we have:
\begin{alignat*}{2}
&D(H(x_1^0)\|H(x_1^1)) \leq D(P^{q/\gamma,n}_{p,\beta}\|Q^{q/\gamma,n}_{p,\beta}),
\end{alignat*}
with $p=+\infty$, $\beta=\lambda_{k}$, and $q=d\cdot \lambda_{k}$.}
\end{lemma}
\begin{IEEEproof}
As the Poisson randomizer involves with a special (empty) output $\Phi$, the original variation-ratio analyses \cite{wang2023unified} is not not applicable (causing an infinity ratio $q$ over $\Phi$). \\

To prove the conclusion, we need to trace back to the intermediate mixture decomposition result in \cite[lemma 4.4]{wang2023unified}: \emph{Given $x_1^0,x_1^1,...,x_n\in \mathbb{X}$, 
if algorithms $\{\mathcal{R}_i\}_{i\in [n]}$ satisfy the $(p, \beta')$-variation property and the $q$-ratio property with some $p> 1, q\geq 1$ and $\beta'=D_{1}(\mathcal{R}_1(x_1^0)\| \mathcal{R}_1(x_1^1))$, then there exists distributions $\mathcal{Q}_1^0, \mathcal{Q}_1^1, \mathcal{Q}_1, \mathcal{Q}_2, ..., \mathcal{Q}_n$ such that
\begin{align}
\label{eq:mix1}
&\mathcal{R}_1(x_1^0) = p \alpha \mathcal{Q}_1^0 + \alpha \mathcal{Q}_1^1+(1-\alpha-p \alpha)\mathcal{Q}_1 \\
\label{eq:mix2}
&\mathcal{R}_1(x_1^1) = \alpha \mathcal{Q}_1^0 + p \alpha  \mathcal{Q}_1^1+(1-\alpha-p\alpha)\mathcal{Q}_1 \\
\label{eq:mix3}
&\forall i\in [2,n],\  \mathcal{R}_i(x_i) = r \mathcal{Q}_1^0 + r \mathcal{Q}_1^1+(1-2r)\mathcal{Q}_i
\end{align}
where $\alpha=\frac{\beta'}{p-1}$ and $r=\frac{\alpha p}{q}$.}

\vspace{\baselineskip}
\noindent
Let $p=+\infty$, $\beta'=\lambda_k$, and $q=d\cdot \lambda_k$, we have the Poisson randomizer $\mathcal{R}_{k}(\cdot)$ satisfies the $(p,\beta')$-variation property, and the uniform blanket message satisfy the $q$-ratio property w.r.t. $\mathcal{R}_{k}(\cdot)$. By letting $\mathcal{Q}_1^0$ as the Dirac delta distribution on $x_1^0$, $\mathcal{Q}_1^1$ as the Dirac delta distribution on $x_1^1$, we can also have the mixture decomposition in Equations \ref{eq:mix1}, \ref{eq:mix2}, and \ref{eq:mix3} holds. Plugin this result into the procedure of variation-ratio reduction \cite{wang2023unified}, we have the final conclusion.
\end{IEEEproof}

\vspace{\baselineskip}
\noindent
Combining Theorem \ref{the:randomreduction}, Lemma \ref{lemma:poissonreduction}, and Lemma \ref{lemma:simplification}, we can simplify the privacy guarantee when each user holds only one element (i.e., $s \equiv 1$) in Algorithm \ref{alg:set} to a formula with respect to $n = n' \cdot \lceil m \rceil$, $\gamma = \frac{m}{\lceil m \rceil}$, $p = +\infty$, $\beta = \lambda_{k}$, and $q = d \cdot \lambda_{k}$.

\subsubsection{Privacy Guarantees}\label{subsec:privacy}
\noindent
The previous privacy guarantee applies to the case where each user contributes (at most) one input-dependent message (i.e., $s \equiv 1$). In this part, we derive the user-level privacy guarantees, taking into account the privacy leakage due to the overall $s$ invocations of the Poisson randomizer by the user.  \\

\noindent
Essentially, user $i$ in lines 9-12 of Algorithm \ref{alg:set} can be seen as $s$ virtual users, each holding only one item. The privacy guarantee of one virtual user contributing one input-dependent message has been established in Lemma \ref{lemma:poissonreduction}, denoted as $(\epsilon', \delta')$-DP. Then, according to the group privacy composition of DP \cite{vadhan2017complexity} (in Lemma \ref{lemma:group}), the group of $s$ virtual users satisfies $(s \cdot \epsilon', s \cdot \exp(s \cdot \epsilon') \cdot \delta')$-DP. Therefore, the user-level privacy guarantee of each \emph{user} in Algorithm \ref{alg:set} is straightforward, as stated in Theorem \ref{the:privacy}.

\begin{theorem}[Privacy Guarantee of Algorithm \ref{alg:set}]\label{the:privacy}
For neighboring dataset $X=\{x_1,\ldots,x_n\}\in \mathbb{X}^n$ and $X'=\{x'_1,\ldots,x'_n\}\in\mathbb{X}^n$ that differ at the $i$-th user's data (with the same privacy levels across users), let $G(X), G(X')$ denote the output of Algorithm \ref{alg:set} given input of $X$ and $X'$, respectively. Let $n'=n\cdot \lceil m \rceil$, $\gamma=\frac{m}{\lceil m \rceil}$, $p=+\infty$, $\beta=\lambda_{k}$, and $q=d\cdot \lambda_{k}$, if $D_{e^{\epsilon'}}(P^{q/\gamma,n'}_{p,\beta}\|Q^{q/\gamma,n'}_{p,\beta})\leq \delta'$, then we have
\begin{alignat*}{2}
&D_{e^\epsilon}(G(X)\|G(X'))\leq \delta,
\end{alignat*}
with $\epsilon=s\cdot \epsilon'$ and $\delta=s\cdot\exp(s\cdot \epsilon')\cdot \delta'$.
\end{theorem}

\vspace{\baselineskip}
\noindent
Actually, our privacy amplification guarantee in Theorem \ref{the:privacy} is almost tight. Specifically, our privacy amplification is exactly tight when $s = 1$ (see the privacy amplification lower bound in Proposition \ref{pro:lower}). However, the group composition of DP uses an approximate (i.e., simplified) formula on the $\delta$ parameter, which causes a minor gap \cite{dwork2006differential}.

\begin{proposition}[Privacy Lower Bound of Algorithm \ref{alg:set}]\label{pro:lower}
Let $n'=n\cdot \lceil m \rceil+1$, $\gamma=\frac{m}{\lceil m \rceil}$, $p=+\infty$, $\beta=\lambda_{k}$, and $q=d\cdot \lambda_{k}$, then there exists neighboring dataset $X=\{x_1,\ldots,x_n\}\in \mathbb{T}^{1\times n}$ and $X'=\{x'_1,\ldots,x'_n\}\in\mathbb{T}^{1\times n}$ that differ at the $i$-th user's data, with  $G(X), G(X')$ denote the output of Algorithm \ref{alg:set} given input of $X$ and $X'$ (and with the same privacy level options $L$) respectively, such that for any distance measure $D$ satisfying the data processing inequality, 
\begin{alignat*}{2}
&D(G(X)\|G(X'))\geq D(P^{q/\gamma,n'}_{p,\beta}\|Q^{q/\gamma,n'}_{p,\beta}).
\end{alignat*}
\end{proposition}
\begin{IEEEproof}
As the privacy level options $L$ is the same for $X$ and $X'$, and the level aggregation results $n_1,\ldots,n_K$ are noisyless, we only need to analyze the counts $C=\{C_1,\ldots,C_d\}$ at line 20 of Algorithm \ref{alg:set}. We construct two datasets $X$ and $X'$ such that:(1) $x_i=\{T_{j_0}\}$ and $x_i=\{T_{j_0}\}$ where $j_0,j_1\in [d]$ and $j_0\neq j_1$; (2) for all $i'\in [n]\backslash\{i\}$, $T_{j_0}\not in x_{i'}$ and $T_{j_1}\not in x_{i'}$ hold. Let $C^0$ and $C^1$ denote the counts $C$ of Algorithm \ref{alg:set} given input data $X$ and $X'$ respectively, to prove the conclusion, it is enough to show $D(C^0\|C^1)\geq D(P^{q/\gamma,n'}_{p,\beta}\|Q^{q/\gamma,n'}_{p,\beta})$. To this end, we define the following post-processing function $F_{counts}$ on the counts $C$:
\begin{itemize}
    \item return $(C_{j_0},C_{j_1})$
\end{itemize}
Then, it is obvious that $F_{counts}(C^0)$ follows the same distribution as $P^{q/\gamma,n'}_{p,\beta}$, and $F_{counts}(C^1)$ follows the same distribution as $Q^{q/\gamma,n'}_{p,\beta}$. Therefore, according to the data processing inequality, we have $D(C^0\|C^1)\geq D(P^{q/\gamma,n'}_{p,\beta}\|Q^{q/\gamma,n'}_{p,\beta})$.
\end{IEEEproof}

\subsection{Optimize Protocol Parameters}\label{subsec:hyperparameter}
\noindent
In this section, we aim to optimize the protocol parameters (at line 6 of Algorithm \ref{alg:set}). To achieve this goal, we first analyze the mean squared error (MSE) of Algorithm \ref{alg:set} with fixed values of $m$ and $\lambda_{1}, \ldots, \lambda_{K}$. We then seek appropriate values for these parameters to approximately minimize the error.

\noindent
\textbf{Utility guarantees.} With fixed parameters $m$ and $\lambda_{1}, \ldots, \lambda_{K}$, the estimation error of Algorithm \ref{alg:set} comes from three sources: (1) the sampling error of the $n$ samples from distribution $P_{\mathbb{T}^s}$; (2) the error due to the Poisson randomizer for input-dependent messages; (3) the error due to blanket messages. Assuming the (unknown) distribution of set-valued data is $P_{\mathbb{X}^s}$, the true element distribution is $W = \{w_1, \ldots, w_d\} = \sum_{x \in \mathbb{X}^s} x \cdot P_{\mathbb{X}^s}(x)$, where $x \in \{0,1\}^d$ is represented in vector form. Each sample $x_i \sim P_{\mathbb{X}^s}$ is an unbiased estimation of the true element distribution. In the binary vector form, the input-dependent signal $x_{i,j} \in \{0,1\}$ follows a Bernoulli distribution $Bernoulli(w_j \cdot \lambda_k)$ with success rate $w_j \cdot \lambda_k$. Therefore, its variance due to sources (1) and (2) is $w_j \cdot \lambda_k (1 - w_j \cdot \lambda_k)$. The bucket $j \in [d]$ of each blanket message follows a Bernoulli distribution $Bernoulli(\frac{m}{d \lceil m \rceil})$, thus its variance is $\frac{m}{d \lceil m \rceil} \cdot \left(1 - \frac{m}{d \lceil m \rceil}\right)$. Iterating over all $i \in [n]$ and $j \in [d]$, we have the overall variance of the occurrences $\{C_j\}_{j \in [d]}$ at line 20 as:
\begin{alignat*}{2}
&\sum_{j\in [d]}\textsf{Var}[C_j]\\
\leq &\sum_{j\in [d]}(\sum_{k\in [K]} n_k\cdot w_j\cdot \lambda_k(1-w_j\cdot \lambda_k)\\
&+{d\cdot n\cdot \lceil m\rceil \cdot \frac{m}{d\lceil m\rceil} \cdot (1-\frac{m}{d\lceil m\rceil})}\\
\leq & n\cdot  m+\sum_{k\in [K]}\sum_{j\in [d]} n_k\cdot w_j\cdot \lambda_k(1-w_j\cdot \lambda_k)\\
\leq & n\cdot  m+\sum_{k\in [K]} n_k\cdot s\cdot \lambda_k.
\end{alignat*}

\vspace{\baselineskip}
\noindent
Assuming the level histogram $\{n_1,\ldots,n_k\}$ are noiseless, the occurrences are debiased at line 21, we then have the mean squared error of the final estimator $\hat{W}$ as:
\begin{alignat*}{2}
&\sum_{j\in [d]}\mathbb{E}[|w_j-\hat{w}_j|^2]\\
\leq& \frac{\sum_{j\in [d]}\textsf{Var}[C_j]}{(\sum_{i\in [n]} n_k\cdot \lambda_{k})^2}\\
\leq & \frac{n\cdot  m+\sum_{k\in [K]}\sum_{j\in [d]} n_k\cdot w_j\cdot \lambda_k(1-w_j\cdot \lambda_k)}{(\sum_{k\in [K]} n_k\cdot \lambda_{k})^2}\\
\leq & \frac{n\cdot  m+s\cdot \sum_{k\in [K]} n_k\cdot \lambda_k}{(\sum_{k\in [K]} n_k\cdot \lambda_{k})^2}.
\end{alignat*}

\vspace{\baselineskip}
\noindent
\textbf{Parameter choices under privacy and communication constraints.} Intuitively, both the number $m$ of blanket messages and the Poisson sampling rate $\lambda_{k}$ affect the privacy guarantee of users with $\epsilon_{i} = E_{k}$, and approximately $n \cdot m$ blanket messages are shared among all users. In this part, we consider the case where the number of blanket messages $m$ is specified. That is, the communication overhead for blanket noises is constrained. Then, the problem of finding optimal hyper-parameters $\lambda_{1}, \ldots, \lambda_{K}$ that can minimize the MSE bound under DP constraints can be formulated as follows:
\begin{alignat*}{2}
\min_{\lambda_{1},\ldots,\lambda_{k}}& \frac{n\cdot  m+s\cdot \sum_{k\in [K]} \hat{n}_k\cdot \lambda_k}{(\sum_{i\in [n]} \hat{n}_k\cdot \lambda_{k})^2}\\
s.t.\ \ & D_{e^{E_k/s}}(P_{+\infty,\lambda_{k}}^{d\cdot \lambda_{k}/\gamma,n'}\| Q_{+\infty,\lambda_{k}}^{d\cdot \lambda_{k}/\gamma,n'})\leq \delta/(s\cdot e^{E_k}),\\
& 0\leq \lambda_{k}\leq 1\text{, for } k\in [K],
\end{alignat*}
where $n'=n\cdot \lceil m\rceil$ and $\gamma=\frac{m}{\lceil m \rceil}$.
As $m$ is fixed, it is obvious that the objective MSE decreases with $\lambda_{1}, \ldots, \lambda_{k}$. Thus, finding the optimal parameters is equivalent to solving the following problem:
\begin{alignat}{2}\label{eq:mlambda}
\arg & \max_{0 \leq \lambda_{k} \leq 1} \lambda_{k} \\
s.t.\ \ & D_{e^{E_k/s}}(P_{+\infty,\lambda_{k}}^{d \cdot \lambda_{k}/\gamma, n'} \| Q_{+\infty, \lambda_{k}}^{d \cdot \lambda_{k}/\gamma, n'}) \leq \delta/(s \cdot e^{E_k}).
\end{alignat}
Since the Hockey-stick divergence grows with $\lambda_{k}$, by performing a binary search over $[0,1]$, we can find an approximately optimal $\lambda_{k}^*$ with $2^{-t}$ additive error within $t$ steps. Each step will cost $\tilde{O}(n')$ time \cite{wang2023unified}.

\noindent
\textbf{Parameter choices under privacy constraints.} The problem of finding optimal hyper-parameters (including the number of blanket messages per user and the Poisson sampling rate) that can minimize the MSE bound under DP constraints is formulated as follows:
\begin{alignat*}{2}
\min_{m,\lambda_{1},\ldots,\lambda_{k}}& \frac{n\cdot  m+s\cdot \sum_{k\in [K]} \hat{n}_k\cdot \lambda_k}{(\sum_{i\in [n]} \hat{n}_k\cdot \lambda_{k})^2}\\
s.t.\ \ &m\in \mathbb{R}^{\geq 0},\\
& D_{e^{E_k/s}}(P_{+\infty,\lambda_{k,}}^{d\cdot \lambda_{k}/\gamma,n'}\| Q_{+\infty,\lambda_{k}}^{d\cdot \lambda_{k}/\gamma,n'})\leq \delta/(s\cdot e^{E_k}),\\
& 0\leq \lambda_{k}\leq 1\text{, for } k\in [K],
\end{alignat*}
where $n'=n\cdot \lceil m\rceil$ and $\gamma=\frac{m}{\lceil m \rceil}$.

\vspace{\baselineskip}
\noindent
\editg{Let $m_k$ denote the minimum value that allows $\lambda_k$ to be the maximum possible value of 1:
\begin{alignat*}{2}
\arg & \min_{m \in \mathbb{R}^{\geq 0}} m \\
s.t.\ & D_{e^{E_k/s}}(P_{+\infty, 1}^{d \cdot \frac{\lceil m \rceil}{m}, n \cdot \lceil m \rceil} \| Q_{+\infty, 1}^{d \cdot \frac{\lceil m \rceil}{m}, n \cdot \lceil m \rceil}) \leq \frac{\delta}{s \cdot e^{E_k}}.
\end{alignat*}
}

\vspace{\baselineskip}
\noindent
Essentially, the hyperparameter $m$ should lie between $(m_1, m_K]$, otherwise the protocol degrades to a uniform-privacy setting (when $m \leq m_1$) or it will unnecessarily introduce sampling error to all users (when $m > m_K$). Since the optimization problem is complicated and involves $\tilde{O}(n')$-complexity Hockey-stick divergence computation, one may numerically search for an optimal $m^*$ in the range $(m_1, m_K]$ to approximately minimize the theoretical MSE using strategies like grid search. After the number of blanket messages is fixed to $m^*$, the optimal parameters $\lambda_{1}, \ldots, \lambda_{K}$ can then be found according to Equation \ref{eq:mlambda}.

\vspace{\baselineskip}
\noindent
Essentially, when $n$ is sufficiently large, according to the asymptotic privacy amplification bound in Equation \ref{eq:nprivacy}, we have $m_1=O\left(\frac{ds^2\log(1/\delta)}{n \cdot E_1^2}\right)$ and $m_K=O\left(\frac{ds^2\log(1/\delta)}{n \cdot E_K^2}\right)$ that decrease with $n$, indicating that our protocol requires not many messages per user and thus is communication efficient.

\section{Experiments}\label{sec:exp}
\noindent
In this section, we evaluate the utility and efficiency of our personalized multi-message shuffle DP protocol.

\noindent
\textbf{Datasets.} We use both synthetic set-valued data and real-world set-valued user data from the MSNBC dataset \cite{misc_msnbc}. In the MSNBC dataset that describes the page visits of users who visited msnbc.com, the item domain size is $T=17$, with each user having at most 17 items and an average of 2 items per user. We simulate scenarios with 5,000 and 50,000 users, each with 4 items (after item padding or random selection). For the synthetic dataset, we simulate a relatively larger item domain size of $T=128$, where each user holds 4 or 8 uniformly random items from the item domain.

\noindent
\textbf{Segmented privacy levels.} To simulate segmented privacy scenarios, we set privacy level options $E=\{0.5, 1.0, 2.0\}$, covering both the conservative case $\epsilon=0.5$ (i.e., privacy adversaries have at most an $e^{0.5}\approx 1.65$ multiplicative posterior gain about the victim's data \cite{dwork2008differential}) and the liberal case $\epsilon=2$ (i.e., privacy adversaries can have at most an $e^2\approx 7.39$ multiplicative probability gain). We assume three segmented settings:
$$S_1=\{25\%\cdot n,\ 50\%\cdot n,\  25\%\cdot n\},$$
$$S_2=\{50\%\cdot n,\ 25\%\cdot n,\ 25\%\cdot n\},$$
$$S_3=\{25\%\cdot n,\ 25\%\cdot n,\ 50\%\cdot n\}.$$ In $S_1$, 25\% of the user population chooses $E_1=0.5$, half users choose $E_2=1$, and the remaining 25\% chooses $E_3=2$. In $L_2$, more users choose the conservative level, while in $L_3$, more users choose the liberal level.

\noindent
\textbf{Competitive approaches.} We compare our protocol to the following approaches in decentralized settings:
\begin{itemize}
    \item \textbf{Subexp, local}: This approach uses the state-of-the-art local DP randomizer: subset exponential mechanism \cite{wang2018privset}, in the local DP model, assuming \emph{all users} adopt the liberal privacy level $\epsilon=2$. This approach does not satisfy segmented/personalized DP, but can be seen as the utility upper bound of personalized LDP approaches (e.g., in \cite{nie2018utility, xue2022mean}).
    \item \textbf{Subexp, shuffle}: Since there is no existing work on set-valued data with personalized DP in the single-message model, this approach uses the subset exponential randomizer in the single-message shuffle model of DP, assuming \emph{all users} adopt the privacy level $\epsilon=0.5$. Privacy is amplified using the SOTA numerical amplification approach of variation-ratio reduction.
    \item \textbf{MM, shuffle}: This approach separately runs the SOTA multi-message shuffle protocol: balls-into-bins \cite{luo2022frequency}, assuming \emph{all users} adopt the privacy level $\epsilon=0.5$.
    \item \textbf{SepMM, shuffle}: This approach separately runs the SOTA multi-message shuffle protocol: balls-into-bins, for each segmentation in $E$. For example, $25\% \cdot n$ users run the balls-into-bins protocol with $(0.5, 0.01/n)$-DP, $50\% \cdot n$ users run the balls-into-bins protocol with $(1, 0.01/n)$-DP, and $25\% \cdot n$ users run the balls-into-bins protocol with $(2, 0.01/n)$-DP. Three estimators from these protocols are averaged as the final estimator.
    \item \textbf{Weighted SepMM, shuffle}: This approach works the same as \emph{SepMM, shuffle}, except the three estimators are weighted averages. Specifically, since the MSE bound of the balls-into-bins protocol \cite{luo2022frequency} with $(\epsilon, \delta)$-DP is $O(ds^2 \log(1/\delta)/(n \cdot \epsilon)^2 + s/n)$, we use the weight $\frac{1}{\sqrt{ds^2 \log(1/\delta)/(n \cdot \epsilon)^2 + s/n}}$ for each estimator to form the weighted average estimator. This is a strong competitor for personalized DP in the shuffle model, as it utilizes users' data from all segmentation and exploits the signal-to-noise information of the three estimators.
\end{itemize}

\begin{figure}[!htb]
\begin{center}
\centerline{\includegraphics[width=80mm]{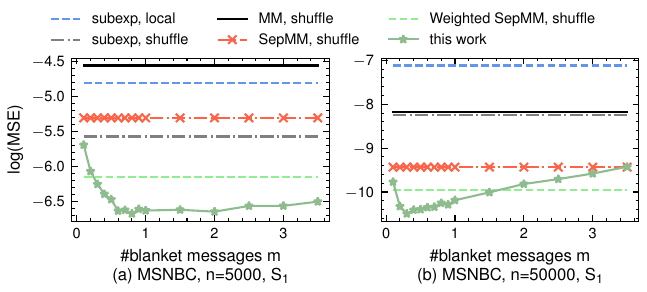}}
\vspace*{-0.8em}
\caption{Experimental results on the MSNBC dataset with the level setting $S_1$.}
\label{fig:MSNBC}
\end{center}
\vspace*{-1.2em}
\end{figure}

\begin{figure}[!htb]
\begin{center}
\centerline{\includegraphics[width=80mm]{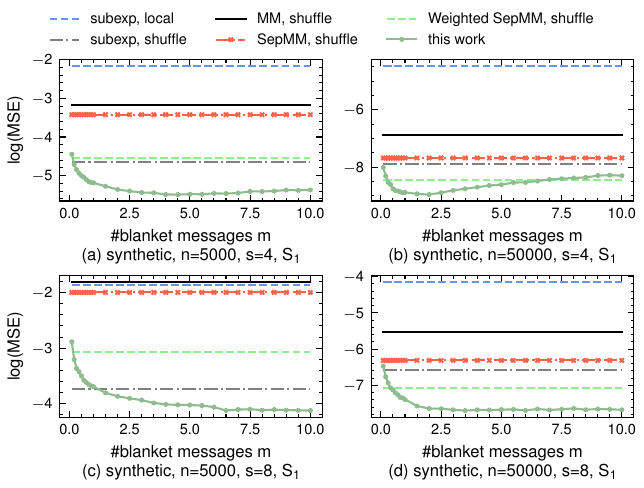}}
\vspace*{-0.8em}
\caption{Experimental results on synthetic dataset with $T=128$, and the level setting $S_1$.}
\label{fig:syn}
\end{center}
\vspace*{-1.2em}
\end{figure}

\noindent
In our experiments, the privacy level options from users are anonymized and shuffled (but not sanitized), the failure probability $\delta$ of DP is set to $0.01/n$. The reported error is mean squared error (MSE): $\sum_{j\in [d]}\|\hat{w}_j-w_j\|_2^2$.

\subsection{Effects of Number of Blanket Messages $m$}
\noindent
In Figures \ref{fig:MSNBC} and \ref{fig:syn}, we present the experimental results with segmentation setting $S_1$ on the MSNBC dataset and the synthetic dataset, respectively. We vary the number of blanket messages per user $m$ from $0.1$ to $10.0$. It is observed that with an appropriate $m$, our protocol achieves much lower MSE than existing approaches, reducing error by about $70\%$. \\

\noindent
On the MSNBC dataset, when the user population size is relatively small (e.g., $n=5,000$), the optimal choice of $m$ is around $2.0$; when the user population size is relatively large (e.g., $n=50,000$), the optimal choice of $m$ is around $0.3$. On the synthetic dataset with $T=128$, $s=4$, and $n=5,000$, the optimal choice of $m$ is around $4.0$; when the user population size is relatively large (e.g., $n=50,000$), the optimal choice of $m$ is around $1.0$. Similar phenomena are observed in cases where $s=8$ in the synthetic dataset. Essentially, if $m$ is too small, every user must inject more noise into the data-dependent message (which is not shared by other users), leading to poor utility performance; if $m$ is too large, the actual privacy guarantee of some (liberal) users might surpass the specified level (meaning more blanket noises are imposed than needed). \\

\noindent
Comparing the cases of the MSNBC dataset (with $T=17$ and $s=4$) and the synthetic datasets with $T=128$ and $s=4$, the optimal choices of $m$ in the synthetic dataset significantly grow. This is possibly because more blanket messages are required to guarantee DP when the domain size is large. Comparing the cases of $s=8$ and $s=4$ in the synthetic datasets, the optimal choices of $m$ for $s=8$ significantly increase because more information should be extracted from users when the original set-valued data contains more items (to avoid large sampling error in the Poisson randomizer).

\subsection{Effects of Number of Users}
\noindent
On the MSNBC dataset, when comparing the scenarios of $n=5,000$ and $n=50,000$, the optimal choice of $m$ decreases as the user population increases. This trend may be attributed to the fact that the total number of blanket messages, $n \cdot m$, approximately dictates the privacy-utility trade-offs. Therefore, as $n$ increase, the $m$ should decrease correspondingly. Notaly, the minimum achievable error for $n=50,000$ is approximatly $1/30$ of that for $n=5,000$, indicating a gap ranging between $1/10$ and $1/10^2$.


\subsection{Effects of Privacy segmentation Settings}
\begin{figure}[!htb]
\begin{center}
\centerline{\includegraphics[width=80mm]{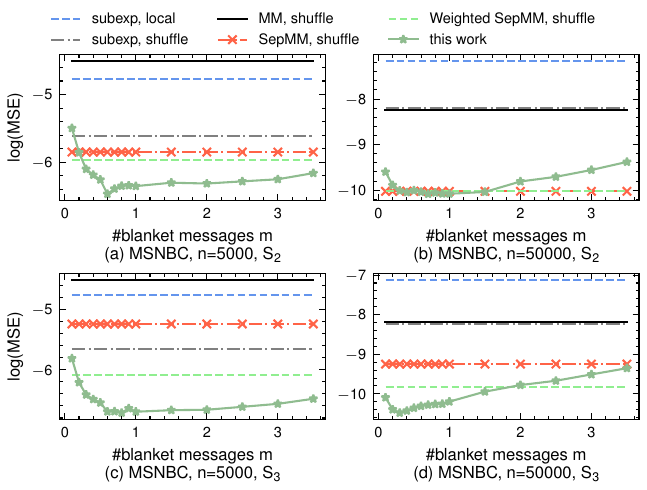}}
\vspace*{-0.8em}
\caption{Experimental results on the MSNBC dataset with the level setting $S_2$ (a,b) and setting $S_3$ (c,d).}
\label{fig:MSNBCS23}
\end{center}
\vspace*{-1.2em}
\end{figure}

\noindent
In Figure \ref{fig:MSNBCS23}, we present the experimental results of the MSNBC dataset with segmentation  settings $S_2=\{50\%\cdot n, 25\%\cdot n, 25\%\cdot n\}$, and 
$S_3=\{25\%\cdot n, 25\%\cdot n, 50\%\cdot n\}$. With more users adopting liberal privacy levels in $S_3$ than $S_2$, the \emph{SepMM, shuffle} approach incurs even more noise (as the estimator from the level $E_0$ introduces much noise), while our approach and \emph{Weighted SepMM, shuffle} will have less error. Additionally, the error gap between our approach and \emph{Weighted SepMM, shuffle} becomes larger in $S_3$. This indicates that our approach can better exploit segmented privacy levels.

\begin{figure}[!htb]
\begin{center}
\centerline{\includegraphics[width=80mm]{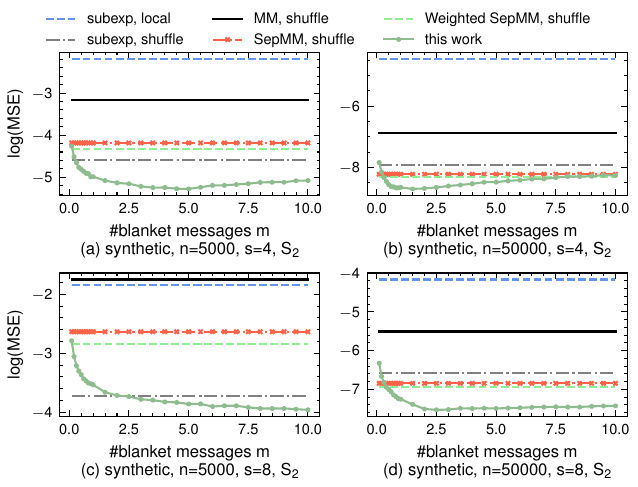}}
\vspace*{-0.8em}
\caption{Experimental results on synthetic dataset with $T=128$, and the level setting $S_2$.}
\label{fig:synS2}
\end{center}
\vspace*{-1.2em}
\end{figure}

\begin{figure}[!htb]
\begin{center}
\centerline{\includegraphics[width=80mm]{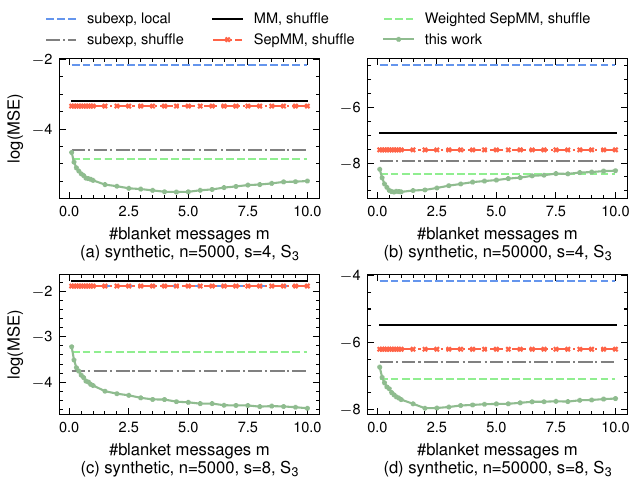}}
\vspace*{-0.8em}
\caption{Experimental results on synthetic dataset with $T=128$, and the level setting $S_3$.}
\label{fig:synS3}
\end{center}
\vspace*{-1.2em}
\end{figure}

\vspace{\baselineskip}
\noindent
In Figures \ref{fig:synS2} and \ref{fig:synS3}, we present the experimental results for the synthetic dataset under segmentation settings $S_2$ and $S_3$, respectively. Notably, with more users adopting liberal privacy levels in $S_3$ than $S_2$,  our approach outperforms existing methods more significantly in $S_3$.

\subsection{Summary}
\noindent
\edit{In both real-world and synthetic datasets, our approach reduces error by about 50\% compared to existing approaches. Compared to single-level approaches in the shuffle model, such as the single-message protocol \emph{Subexp, shuffle} and the multi-message protocol \emph{MM, shuffle}, our approach allows liberal users to inject less noise, thereby introducing less overall noise. Additionally, compared to separated aggregation approaches for each user segmentation in the multi-message shuffle model, such as \emph{SepMM, shuffle} and \emph{Weighted SepMM, shuffle}, our method enables all users to share the noise required for differential privacy, thus reducing the total noise introduced. Additionally, the optimal number $m^*$ of blanket messages in our approach is often below 4, and $m^*$ decreases with $n$, demonstrating excellent communication efficiency.}

\section{Conclusion}\label{sec:conclusion}
\noindent
This work introduces a segmented differentially private aggregation framework within the (multi-message) shuffle model, benefiting both users and analysts by safeguarding user data and privacy levels while leveraging the superior privacy-utility trade-offs of the multi-message shuffle model. We have developed a concrete protocol for set-valued data analyses under this framework and have established almost tight privacy amplification bounds. In addition to theoretically bounding the error of the protocol, we have conducted evaluations on both real-world and synthetic datasets. The results confirm the exceptional utility and efficiency of our approach, which halves estimation errors and requires only a few messages per user.

\edit{\noindent\textbf{Discussion on Practicality} While achieving excellent segmented privacy-utility trade-offs, our framework and concrete protocols are also efficient. It requires only two rounds of interactions between each user and the shuffler/server, and each user contributes only a few messages (about four messages, see Section \ref{sec:exp}). Since users’ privacy levels are segmented (instead of being fully personalized), the additional Step 2 (Optimizing protocol parameters) of our protocol on the server side is also efficient.}


\renewcommand\refname{\zihao{5}\textbf{References}}

\bibliographystyle{unsrt}
\bibliography{refsall}

\end{document}